\documentclass{article}
\usepackage{spconf,amsmath,graphicx}
\usepackage{times}
\usepackage{epsfig}
\usepackage{graphicx}
\usepackage{amsmath}
\usepackage{amssymb}
\usepackage{algorithm}
\usepackage{algorithmic}
\usepackage{enumerate}
\usepackage{multirow}
\usepackage{multicol}
\usepackage{arydshln}
\usepackage{color}
\usepackage{amssymb}
\def\0{{\mathbf 0}}
\def\1{{\mathbf 1}}

\def\e{{\mathbf e}}

\def\g{{\mathbf g}}

\def\l{{\mathbf l}}

\def\p{{\mathbf p}}

\def\r{{\mathbf r}}

\def\v{{\mathbf v}}

\def\x{{\mathbf x}}
\def\y{{\mathbf y}}
\def\z{{\mathbf z}}

\def\B{{\mathbf B}}

\def\D{{\mathbf D}}

\def\F{{\mathbf F}}
\def\G{{\mathbf G}}
\def\F{{\mathbf F}}
\def\H{{\mathbf H}}
\def\I{{\mathbf I}}

\def\L{{\mathbf L}}

\def\W{{\mathbf W}}
\def\X{{\mathbf X}}

\def\ie{{\textit{i.e.}}}
\def\eg{{\textit{e.g.}}}

\def\cG{{\mathcal G}}

\def\cK{{\mathcal K}}
\def\cL{{\mathcal L}}

\def\cN{{\mathcal N}}
\def\cO{{\mathcal O}}

\def\cU{{\mathcal U}}

\DeclareMathOperator{\exE}{\mathbb{E}}

\title{Fast \& Robust Image Interpolation using Gradient Graph Laplacian Regularizer}
\name{Fei Chen$^\star$, Gene Cheung$^\dag$, Xue Zhang$^\dag$\thanks{This work was supported by the National Natural Science Foundation of China (61771141) and by China Scholarship Council (CSC).}}
\address{$^\star$College of Mathematics and Computer Science, Fuzhou University,
Fuzhou, China\\
 $^\dag$Dept of EECS, York University, Toronto, Canada}
\begin{document}
\ninept
\maketitle
\begin{abstract}
In the graph signal processing (GSP) literature, it has been shown that signal-dependent graph Laplacian regularizer (GLR) can efficiently promote piecewise constant (PWC) signal reconstruction for various image restoration tasks. 
However, for planar image patches, like total variation (TV), GLR may suffer from the well-known ``staircase" effect. 
To remedy this problem, we generalize GLR to gradient graph Laplacian regularizer (GGLR) that provably promotes piecewise planar (PWP) signal reconstruction for the image interpolation problem---a 2D grid with random missing pixels that requires completion.
Specifically, we first construct two higher-order gradient graphs to connect local horizontal and vertical gradients. 
Each local gradient is estimated using structure tensor, which is robust using known pixels in a small neighborhood, mitigating the problem of larger noise variance when computing gradient of gradients.
Moreover, unlike total generalized variation (TGV), GGLR retains the quadratic form of GLR, leading to an unconstrained quadratic programming (QP) problem per iteration that can be solved quickly using conjugate gradient (CG).
We derive the means-square-error minimizing weight parameter for GGLR, trading off bias and variance of the signal estimate.
Experiments show that GGLR outperformed competing schemes in interpolation quality for severely damaged images at a reduced complexity. 
\end{abstract}
\section{Introduction}
\label{sec:intro}
Due to limitations in image sensing, compression artifacts and transmission errors, acquired images are often imperfect with distorted or missing pixels. 
\textit{Image restoration} is the task of recovering a pristine image from corrupted and/or partial observations.
Appropriate signal priors are required to regularize an otherwise under-determined inverse problem. 
Classical methods like \textit{total variation} (TV) \cite{chambolle1997image} and sparse representation \cite{donoho2003optimally} take a model-based approach, where mathematical signal assumptions are made before data observations.
In contrast, development in \textit{deep learning} (DL) has led to purely data-driven regularizers that perform well on average \cite{zhang2017image, bigdeli2017deep}.
However, pure DL schemes require large training data and memory footprints, and perform poorly when statistics of training and testing data are mismatched \cite{peng2016fine, touvron2019fixing, vu21}.
In this paper, we propose a new model-based prior for image restoration that is both fast and robust. 
We leave the investigation of hybrid schemes that combine advantages of model-based and data-driven approaches \cite{zeng2019deep, su2020graph, vu21} as future work.

Recent interest in \textit{graph signal processing} (GSP) \cite{ortega2018ieee, gene2018ieee}---analysis of discrete signals residing on irregular data kernels described by graphs---has resulted in graph spectral image restoration algorithms for denoising, deblurring, contrast enhancement, etc \cite{Pang2017, bai19, liu19}. 
Extending previous model-based approaches, GSP-based schemes interpret an image $\x$ as a signal on a suitably defined graph $\cG$, and assume $\x$ is smooth or bandlimited with respect to (w.r.t.) $\cG$.  
In particular, the \textit{signal-dependent graph Laplacian regularizer} (SDGLR or GLR for short) \cite{Pang2017,liu17} quantifies smoothness of graph signal $\x$ w.r.t. $\cG$, specified by graph Laplacian matrix $\L$:
\begin{align}
\x^{\top} \L(\x) \x &= 
\sum_{ij} w_{i,j}(x_i,x_j) \, (x_i - x_j)^2 
\label{eq:glr} \\
w_{i,j}(x_i,x_j) &= 
\exp \left( 
- \frac{\| \l_i-\l_j \|^2_2}{\sigma_l^2}
- \frac{|x_i-x_j|^2}{\sigma_x^2} \right)
\label{eq:edgeWeight0}
\end{align}
where $w_{i,j}$ is the weight of an edge connecting nodes $i$ and $j$, and $\l_i$ and $x_i$ are the pixel location and intensity, respectively. 
Note that $w_{i,j}$ is \textit{signal-dependent}---it is a Gaussian function of intensity difference $|x_i - x_j|$---and hence $\L(\x)$ is a function of sought signal $\x$.
This means that each term in the sum in \eqref{eq:glr} is minimized when $|x_i - x_j|$ is either very small (thus $(x_i-x_j)^2$ is small) or very large (thus $w_{i,j}$ is small).
As a result, minimizing GLR promotes \textit{piecewise constant} (PWC) signal reconstruction as proven in \cite{Pang2017,liu17}.

However, like TV that also promotes PWC signal reconstruction, GLR \eqref{eq:glr} suffers from the well-known ``staircase" effect for image patches with linearly changing intensity. 
Extending GLR, in this paper we propose a higher-order graph smoothness prior called \textit{gradient graph Laplacian regularizer} (GGLR) for the \textit{image interpolation} problem---a 2D grid has randomly missing pixels that require completion.  
Specifically, for a target image, we first construct two \textit{gradient graphs} to  connect local horizontal and  vertical gradients. 
Each local gradient is estimated  using \textit{structure tensor} \cite{knutsson2011representing}, which is robust using known pixels in a small neighborhood, thus alleviating the pitfall of large noise variance when computing gradient of gradients.
Unlike GLR, we prove that GGLR promotes a more general \textit{piecewise planar} (PWP) signal reconstruction. 

Computationally, GGLR retains the simple quadratic form in \eqref{eq:glr}, leading to an unconstrained \textit{quadratic programming} (QP) problem per iteration when edge weights are fixed, solvable using fast numerical linear algebra methods like \textit{conjugate gradient} (CG) \cite{axelsson1986rate}. 
Leveraging \cite{chen2017bias}, we derive the \textit{means-square-error} (MSE) minimizing weight parameter for GGLR, trading off bias and variance of the signal estimate.
Experiments show that GGLR is both fast and robust, outperforming competitors in interpolation quality when a large number of pixels are missing at a reduced complexity.

\vspace{0.05in}
\noindent
\textbf{Related Work}: 
Low-complexity classical interpolation methods such as bilinear interpolation \cite{pressnumerical} are commonly used in consumer software, but are sub-optimal in general due to the lack of signal adaptivity in local neighborhoods. 
Edge-guided methods such as \textit{partial differential equations} (PDE) \cite{folland1995introduction} provide smooth interpolation, but perform poorly when missing pixels are considerable. 

TV \cite{chambolle1997image} was a popular image prior due to its simplicity in definition and available optimization algorithms in minimizing convex but non-differentiable $\ell_1$-norm, which has no closed-form solutions. 
Its generalization, \textit{total generalized variation} (TGV) \cite{bredies2010total, bredies2015tgv}, better handles the aforementioned staircase effect, but retains the non-differentiable $\ell_1$-norm. 
Like TGV, GGLR also promotes PWP signal reconstruction, but can be optimized efficiently as an unconstrained $\ell_2$-norm QP problem per iteration---this is the focus of our paper.

\vspace{-0.05in}
\section{Graph construction}
\label{sec:sectionII}

\subsection{Horizontal Gradient Graph}

Given an image $\X \in \mathbb{R}^{M\times N}$, we compute its \textit{horizontal gradient} $\G^h \in \mathbb{R}^{M\times (N-1)}$ as the intensity differences between horizontally neighboring pixels, \ie, 
\begin{align}
G^h_{k,l} = X_{k,l+1} - X_{k,l}   
\end{align}
for $1 \leq k \leq M$ and $1 \leq l \leq N-1$. We rewrite $\G^h$ in vector form $\mathbf{g}^h={\rm vec}(\G^h)\in\mathbb{R}^{M(N-1)\times 1}$, where ${\rm vec}(\cdot)$ denotes the vectorization operator stacking columns of a matrix into a vector. 
Thus we have a matrix representation $\g^h = \F^h \x$, where $\x={\rm vec}(\X)$ and $\F^h \in \mathbb{R}^{M(N-1) \times MN}$ is the horizontal gradient operator.

Next, we define a \textit{horizontal gradient graph} $\cG^h$ consisting of $M(N-1)$ nodes corresponding to elements of $\g^h$. 
Similar to \eqref{eq:edgeWeight0}, we define the weight $w^h_{i,j}$ for an edge connecting nodes $i$ and $j$ as

\vspace{-0.15in}
\begin{small}
\begin{align} 
w^h_{i,j}= \left\{ \begin{array}{ll}
\exp\left(- \frac{|g^h_i - g^h_j|^2}{\sigma^2}\right) & j \in \cN^h_i \\
0 & \mbox{o.w.}  
\end{array}\right.
\label{eq:edgeWeight_h}
\end{align}
\end{small}\noindent
where $\cN_i^h$ is a defined local neighborhood centered at gradient $i$. 
$\sigma > 0$ is a chosen parameter so that edge weights $w^h_{i,j}$ are well distributed between $0$ and $1$. 
Define next an \textit{adjacency matrix} $\W^h \in \mathbb{R}^{M(N-1) \times M(N-1)}$ where $W^h_{i,j} = w^h_{i,j}$.
A graph $\cG^h$ is completely characterized by the adjacency matrix $\W^h$.

\subsubsection{GGLR for Horizontal Gradients}

We next define the diagonal \textit{degree matrix} $\D^h$ with diagonal elements $D^h_{i,i} = \sum_j W^h_{i,j}$. 
We then define a \textit{graph Laplacian matrix} $\L^h \in \mathbb{R}^{M(N-1) \times M(N-1)}$ as $\L^h = \D^h - \W^h \in \mathbb{R}^{M(N-1)\times M(N-1)}$. 
Given that the edge weights are non-negative in \eqref{eq:edgeWeight_h}, one can prove that Laplacian $\L^h$ is \textit{positive semi-definite} (PSD) via the Gershgorin circle theorem \cite{gene2018ieee}. 
Since edge weights are defined w.r.t. gradients $g^h_i$ computed from signal $\x$, Laplacian $\L^h(\x)$ is a function of $\x$. 
We now write GLR for gradient $\g^h$
\begin{align} 
\left(\g^h \right)^{\top} \L^h \g^h &= \sum_{i=1}^{M(N-1)} \sum_{j=1}^{M(N-1)} w^h_{i,j} (g^h_j - g^h_i)^2 
\label{eq:GGLR_H1} \\
&= \x^{\top} \underbrace{( \F^h )^{\top} \L^h \F^h}_{\cL^h} \x
= \x^{\top} \cL^h \x
\label{eq:GGLR_H}
\end{align}
where $\cL^h$ is also PSD since $\L^h$ is PSD.
Thus, GLR for gradient $\g^h$ can also be computed in the pixel domain using $\x$ in quadratic form via matrix $\cL^h$. 
We call \eqref{eq:GGLR_H} the \textit{horizontal GGLR} for signal $\x$.

As an example, consider a 3-pixel row $\x = [x_1 ~~x_2 ~~x_3]^{\top}$. 
Using $\F^h = [1 ~~-1 ~~0; ~~0 ~~1 ~~-1]$ results in gradient $\g^h = \F^h \x = [x_1 - x_2 ~~x_2 - x_3]^{\top} = [g_1 ~~g_2]^{\top}$.
Assuming $w_{i,j}^h = 1, \forall (i,j)$, for simplicity, the corresponding matrices $\L^h$ and $\cL^h$ are:
\begin{align}
\L^h = \left[ \begin{array}{cc}
1 & -1 \\
-1 & 1
\end{array} \right], 
~~
\cL^h = \left[ \begin{array}{ccc}
1 & -2 & 1 \\
-2 & 4 & -2 \\
1 & -2 & 1
\end{array} \right] .
\end{align}
Observe that $\cL^h$ is a Laplacian corresponding to a \textit{signed graph} $\cG$, where $w_{1,2} = w_{2,3} = 2$ and $w_{1,3} = -1$. 
See Fig.\;\ref{fig:imgEx}(a) for an illustration. 
Unlike previous signed graphs in GSP, $\cG$ is not balanced in general \cite{yang21} and has no self-loops \cite{su17}, yet Laplacian $\cL^h$ is guaranteed to be PSD by definition \eqref{eq:GGLR_H}, without any eigenvalue shift \cite{cheung18tsipn}. 
Thus, GGLR constitutes a new usage of signed graphs in GSP.

\subsection{Vertical Gradient Graph}

Similarly, we define the gradient in the vertical direction  $\G^v \in \mathbb{R}^{(M-1)\times N}$ as  $G^v_{k,l} = X_{k+1,l} - X_{k,l}$ for $1 \leq k \leq M-1$ and $1 \leq l \leq N-1$. 
We have a similar matrix representation in the vertical direction $\g^v = \F^v \x$, where  $\F^v \in \mathbb{R}^{(M-1)N \times MN}$ is the vertical gradient operator. 
The \textit{vertical GGLR} is thus $\x^{\top} \cL^v \x$, where $\cL^v = (\F^v)^{\top} \L^v \F^v$.
$\cL^v$ is also PSD since $\L^v$ is PSD. 



\vspace{-0.05in}
\section{Gradient graph Laplacian regularizer}
\label{sec:sectionIII}
\begin{figure}[t]
\begin{center}
\includegraphics[width=0.4\linewidth]{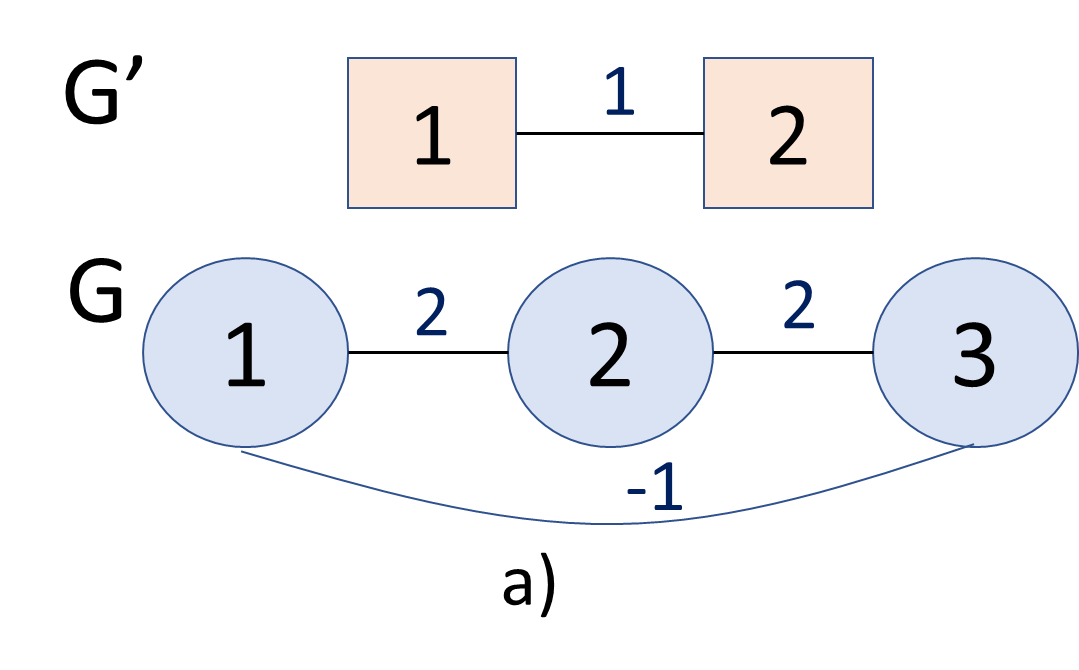}
\includegraphics[width=0.58\linewidth]{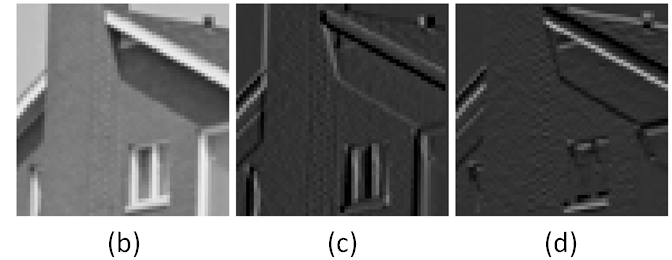}
\end{center}
\vspace{-0.3in}
\caption{(a) Example of 2-node gradient graph $\cG'$ and 3-node pixel graph $\cG$. (b) Image $\mathbf{X}$. (c-d) Horizontal and vertical gradients, $\mathbf{G}^h$  and $\mathbf{G}^v$ respectively, which are approximately piecewise constant.
} 
\label{fig:imgEx}
\end{figure} 

As shown in Fig.\;\ref{fig:imgEx}(b)-(d), natural images are often approximately PWP. 
To promote PWP reconstruction, we define GGLR $\Phi(\x)$ for an image $\x$ by combining the horizontal and vertical GGLRs as
\begin{align}
\Phi(\x) &= \x^{\top} \cL^h \x + \x^{\top} \cL^v \x
= \x^{\top} \left( \cL^h + \cL^v \right) \x .
\label{eq:GGLR}
\end{align}
We first show GGLR promotes PWP signal reconstruction in 1D.


\subsection{Piecewise Planar Signal Reconstruction}
\label{subsec:PWP}

For simplicity, consider a row of $N$ pixels and a corresponding line graph with graph Laplacian $\cL^h = (\F^h)^{\top} \L^h \F^h$.
Gradient operator $\F^h$ in this 1D case is a $N-1$-by-$N$ Toeplitz matrix:
\begin{align}
\F^h = \left[ \begin{array}{cccc}
1 & -1 & 0 & \ldots \\
0 & 1 & -1 & 0 \ldots \\
\vdots & \ddots & \ddots & \ddots
\end{array}
\right] .
\end{align}
We see that the constant vector $\1$ is in the null space of $\F^h$, \ie, $\F^h \1 = \0$. 
This means $\v_1 = \1/\sqrt{N}$ is an eigenvector of $\cL^h$ corresponding to eigenvalue $0$, \ie, $\cL \v_1 = (\F^h)^{\top} \L^h \F^h \v_1 = \0$. 

Consider next a linearly decreasing vector $\x$'s, \ie, 
\begin{align}
x_{i+1} = x_i - \delta, 
~~~~~
1 \leq i \leq N-1
\label{eq:linear}
\end{align}
where $\delta > 0$ is a constant.
Here, $\F^h \x = \delta \1$, and thus $\cL^h \x = (\F^h)^{\top} \L^h \F^h \x = \delta (\F^h)^{\top} \L^h \1 = \0$, since a constant vector is the first eigenvector for graph Laplacian $\L^h$ corresponding to the smallest eigenvalue $0$ \cite{ortega2018ieee}.
Define $\v_2 = \x/\|\x\|_2$. 
Thus, \textit{any} linear combination $\y = \alpha_1 \v_1 + \alpha_2 \v_2$ for constants $\alpha_i$ (\ie, linear signal of any slope) will also evaluate to $\cL^h \y = \0$. 
Thus, GGLR $\x^{\top} \cL^h \x$, first and foremost, promotes linear signal reconstruction. 

Consider next a $N$-pixel row consists of two linear pieces, of lengths $K$ and $N-K$ and slopes $\delta_1$ and $\delta_2$ respectively, separated by a large $\Delta$
\begin{align}
x_{i+1} = \left\{ \begin{array}{ll}
x_i - \delta_1 & \mbox{if}~~ 1 \leq i \leq K-1  \\
x_i - \Delta & \mbox{if}~~ i = K \\
x_i - \delta_2 & \mbox{if}~~ K+1 \leq i \leq N
\end{array} \right.
\end{align}
where $|\Delta| \gg |\delta_1|, |\delta_2|$.
Neighboring $\g^h_i$'s within each linear piece will compute to the same value, and thus $|\g^h_{i+1} - \g^h_i| = 0$. Adjacent $\g^h_i$'s across the piece boundary will compute to $|\g^h_{i+1} - \g^h_i| = |\Delta - \delta_i| \gg 0$.
By edge weight definition \eqref{eq:edgeWeight_h}, weight $w^h_{i+1,i} \approx 0$, and thus the corresponding term in \eqref{eq:GGLR_H1} is also $\approx 0$. 
Thus, we conclude that signal $\x$ that minimizes GGLR for 1D is linear or piecewise linear.

\vspace{-0.05in}
\section{Image Interpolation using GGLR}
\label{sec:sectionIV}
\vspace{-0.05in}
\subsection{Gradient Estimation using Structure Tensor}

Computation of edge weights in the gradient graph using \eqref{eq:edgeWeight_h} requires the difference of gradients $g^h_i - g^h_j = x_{i+1} - x_i - x_{j+1} + x_j$. 
If each pixel $x_i$ is corrupted by zero-mean additive noise with variance $\sigma_n^2$, then variable $g^h_i - g^h_j$ suffers from \textit{four times} the noise variance $4\sigma_n^2$.
Hence, a robust estimate of gradient $g^h_i$ is essential. 

For the targeted image interpolation problem, we compute local gradients $\g^h_i$ reliably using \textit{structure tensor} \cite{knutsson2011representing}. 
Structure tensor $S(\p)$ at a 2D location $\p$ is a $2 \times 2$ matrix computed using observable horizontal and vertical gradients, $g^h_{\p + \r}$ and $g^v_{\p + \r}$ in a window, \ie, 
\begin{align}
S(\p) &= \left[ \begin{array}{cc}
\sum_{\r \in \cU} \frac{1}{|\cU|} (g^h_{\p + \r})^2     & \sum_{\r \in \cU} \frac{1}{|\cU|} g^h_{\p + \r} g^v_{\p + \r} \\
\sum_{\r \in \cU} \frac{1}{|\cU|} g^h_{\p + \r} g^v_{\p + \r}     & \sum_{\r \in \cU} \frac{1}{|\cU|} (g^v_{\p + \r})^2
\end{array} \right]
\label{eq:ST}
\end{align}
where $\cU$ is a set of displacement vectors for the chosen window (\eg, $\cU = \{ (0,1), (0,-1), (1,0), (-1,0)\}$ for a window containing displacement vectors to the four closest neighbors on a 2D grid). 
We discount any gradient terms $g^h_{\p+\r}$ in the sums in \eqref{eq:ST} that cannot be computed due to missing pixel(s). 

\begin{figure}[t]
\begin{center}
\includegraphics[width=0.6\linewidth]{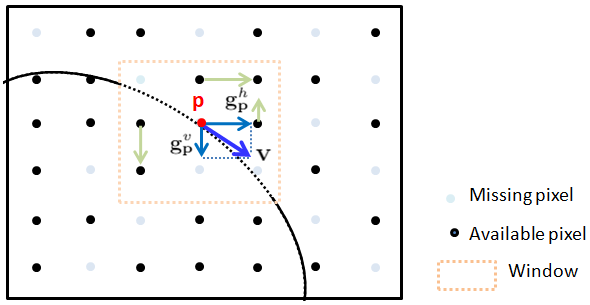}
\end{center}
\vspace{-0.3in}
\caption{Gradient estimation of patch centered at $\p$ using structure tensor $S(\p)$ computed in a window. 
Only gradients $\g^h_{\p+\r}$ and $\g^v_{\p+\r}$ computed using available pixels are used to compute $S(\p)$. } 
\label{fig:ST}
\end{figure}

Upon obtaining $S(\p)$ in \eqref{eq:ST}, we compute the last eigenvector $\e$ corresponding to the larger of two eigenvalues.
$\e$ is the \textit{dominant gradient} of a patch center at $\p$. 
We then project $\e$ into its $x$- and $y$-components as estimates of horizontal and vertical gradients,  $g^h_{\p}$ and $g^v_{\p}$, at $\p$. 
See Fig.\;\ref{fig:ST} for an illustration. 


\subsection{Problem Formulation and Optimization} 

We now formulate the image interpolation problem using GGLR \eqref{eq:GGLR}. 
Denote by $\y \in \mathbb{R}^K$ an observation vector containing the $K$ known pixels with index set $\cK$ in an $M \times N$ image $\X$. 
Denote by $\H \in \{0,1\}^{K \times MN}$ a \textit{selection matrix} that picks out the $K$ known pixels from vectorized $\x = \text{vec}(\X) \in \mathbb{R}^{MN}$, \ie,
\begin{align}
\H_{i,j} = \left\{ \begin{array}{ll}
1 & \mbox{if $j$ is the $i$-th available index in $\cK$} \\
0 & \mbox{o.w.} 
\end{array}
\right. .
\end{align}
The optimal patch $\x^*$ should respect observation $\y$ while minimizes GGLR.
Hence, the optimization objective is
\begin{align} 
\min_{\x}   
\| \H \x - \y \|_{2}^{2} + 
\mu \, \x^{\top} (\cL^h + \cL^v) \x
\label{eq:obj}
\end{align}
where $\mu > 0$ is a parameter trading off the fidelity term and the GGLR prior.
Note that $\cL^h$ and $\cL^v$ are signal-dependent functions of $\x$ due to \eqref{eq:edgeWeight_h}.
As done in \cite{Pang2017,liu17} for GLR, we solve \eqref{eq:obj} via an iterative approach.
For fixed $\cL^h$ and $\cL^v$, we first compute an optimal $\x^*$.
We then update edge weights using \eqref{eq:edgeWeight_h} in the two gradient graphs $\cG^h$ and $\cG^v$ having computed gradients $g^h_i$ and $g^v_i$ from solution $\x^*$.  
We repeat these two steps until convergence. 

For fixed $\cL^h$ and $\cL^v$, \eqref{eq:obj} is an unconstrained QP problem with a closed-form solution: 
\begin{align}
\left( \H^{\top} \H + \mu (\cL^h + \cL^v) \right) \x^* = \H^{\top} \y .
\label{eq:LS}
\end{align}
Coefficient matrix $\B = \H^{\top} \H + \mu (\cL^h + \cL^v)$ is \textit{positive definite} (PD) for sufficiently large $K$. 
For simplicity, we prove only the 1D case when $\x$ is a row of pixels, and $\B$ is PD when $K>1$.  
As shown in Section\;\ref{subsec:PWP}, $\y^{\top} \cL^h \y = 0$ implies that $\y$ is a linear signal.
$\H^{\top} \H$ is a diagonal matrix with 1's and 0's along its diagonal. 
Hence, for $K > 1$, quadratic term for matrix $\H^{\top} \H$ is $\y^{\top} \H^{\top} \H \y = \sum_{i \in \cK} y_i^2 > 0$, since linear signal $\y \neq \0$ can have at most one entry $y_i=0$.
Thus, quadratic terms for $\H^{\top} \H$ and $\cL^h$ cannot both evaluate to $0$ using $\y$.
Given both $\H^{\top} \H$ and $\cL^h$ are PSD, we conclude that $\x^{\top} (\H^{\top} \H + \cL^h) \x > 0, \forall \x \neq \0$, and therefore $\H^{\top} \H + \cL^h$ is PD.
Similarly, $\H^{\top} \H + \mu \cL^h$ is PD for $\mu > 0$.

Given $\B$ is symmetric, sparse and PD, \eqref{eq:LS} can be solved efficiently using \textit{conjugate gradient} (CG) \cite{axelsson1986rate}, without computing matrix inverse, in $\cO(cMN)$, where $c$ is the iteration number in CG.

\subsection{Choosing Tradeoff Parameter $\mu$}
\label{subsec:tradeoff}

Parameter $\mu$ in \eqref{eq:obj} must be carefully chosen for best performance.
Analysis of $\mu$ for GLR-based signal denoising \cite{chen2017bias} showed that the best $\mu$ minimizes the \textit{mean square error} (MSE) by optimally trading off bias of the estimate with its variance. 
Here, we derive a new variant from Corollary 1 in \cite{chen2017bias} to compute $\mu$ for \eqref{eq:obj}. 

For simplicity, consider again a row of pixels where $K = NM$. 
Thus $\H = \I$, and $\B = \I + \mu \cL^h$. 
Assume that observation $\y = \x^o + \z$ is corrupted by zero-mean iid noise $\z$ with covariance matrix $\sigma^2_o \I$.
Denote by $(\lambda_i,\v_i)$ the $i$-th eigen-pair of matrix $\cL^h$.
Since $\lambda_1 = \lambda_2 = 0$, Theorem 2 for MSE in \cite{chen2017bias} can be restated as
\begin{align}
\text{MSE}(\mu) = \sum_{i=3}^K q_i^2(\v_i^{\top} \bar{\x^o})^2 + \sigma^2_o \sum_{i=1}^K h_i^2 
\label{eq:MSE}
\end{align}
where $\bar{\x^o} = \x^o - \frac{\1^{\top} x^o}{K} \1$, $q_i = \frac{1}{1 + \frac{1}{\mu \lambda_i}}$, and $h_i = \frac{1}{1 + \mu \lambda_i}$. The two terms in \eqref{eq:MSE} correspond to bias square and variance of estimate $\x^*$.

Suppose now that ground truth $\x^o = \x^o_p + \z_p$ is a linear signal $\x^o_p$ defined in \eqref{eq:linear} plus zero-mean iid perturbation $\z_p$ with covariance matrix $\sigma_p^2 \I$. 
If gradient graph $\cG^h$ is computed from $\x_p^o$, then $\v_1 = \x_p^o$.
We compute expectation $\exE[(\v_i^{\top} \bar{\x^o})^2]$ for $i \geq 3$:
\begin{align}
\exE[(\v_i^{\top} \x^o)^2] = 
\text{tr} \left(\v_i \v_i^{\top} (\x_p^o (\x_p^{o})^{\top} + \sigma_p^2 \I) \right) = \sigma_p^2 .
\end{align}
We now restate Corollary 1 in \cite{chen2017bias} as an upper bound for MSE:
\begin{align}
\text{MSE}(\mu) \leq 
\frac{(K-2) \sigma_p^2}{\left(1 + \frac{1}{\mu \lambda_K}\right)^2} +
\left(
\frac{K-2}{(1+\mu \lambda_3)^2} + 2
\right) \sigma^2_o .
\label{eq:MSE_bound}
\end{align}
The MSE bound \eqref{eq:MSE_bound} can be minimized by taking the derivative w.r.t. $\mu$ and setting it to 0. 

\vspace{0.2in}

\begin{figure}[t]
\begin{center}
   \includegraphics[width=0.99\linewidth]{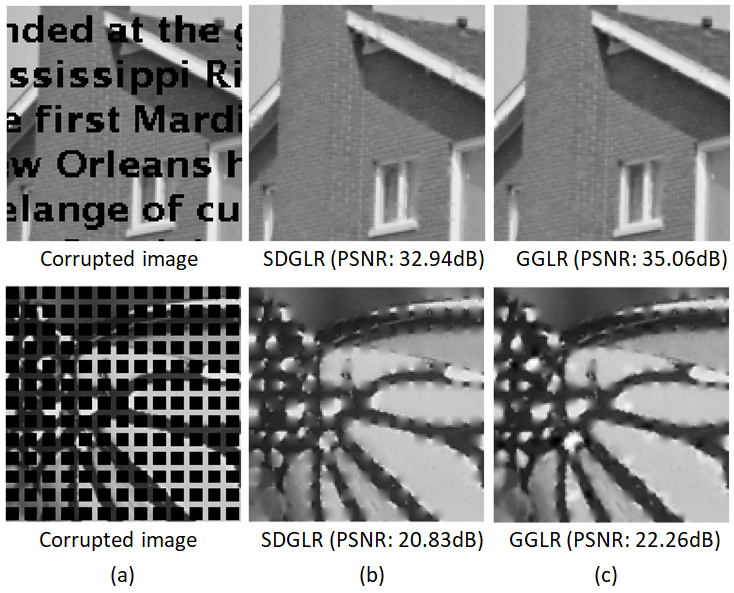}
\end{center}
\vspace{-0.3in}
\caption{Visual and PSNR comparison for patterned missing pixels.  (a) Corrupted images. (b-c) Interpolation by SDGLR and GGLR.
} 
\label{fig:visualResult1}
\end{figure}

\vspace{-0.3in}
\section{Experiments}
\label{sec:results}
\begin{figure}[t]
\begin{center}
   \includegraphics[width=0.99\linewidth]{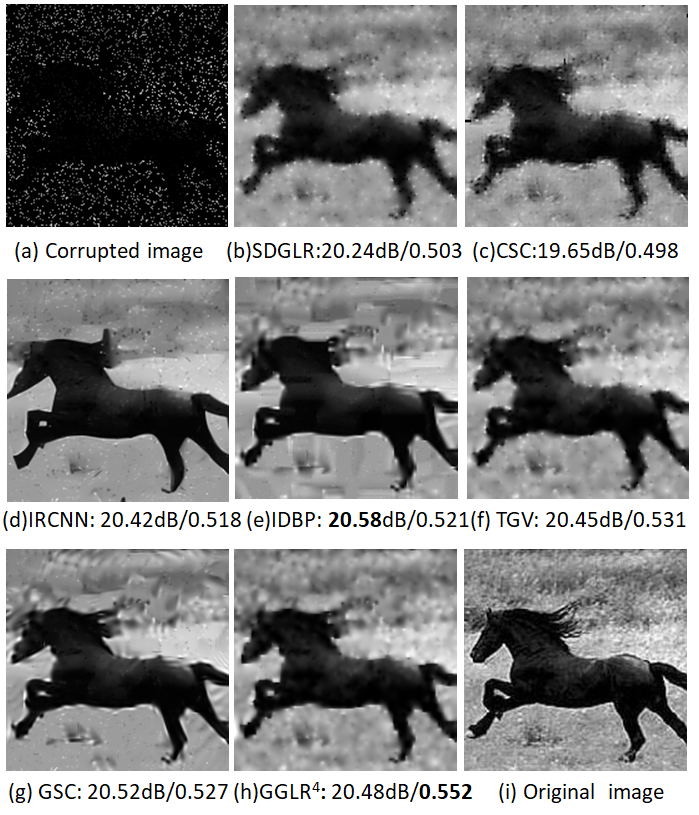}
\end{center}
\vspace{-0.3in}
\caption{Visual, PSNR and SSIM comparison for 90\% missing pixels. Interpolation by GGLR looks less blocky and more natural.} 
\label{fig:visualResult2}
\end{figure}

\vspace{-0.1in}
We first describe parameter setting in our experiments. 
$\cN_i$ was set to 2-neighborhood (4-neighborhood), for 2-connected ( 4-connected) gradient graphs. 
$\sigma$ in \eqref{eq:edgeWeight_h} was set to 0.68. 
Window $\cU$ in \eqref{eq:ST} was set to $5\times5$. 
Based on MSE analysis in Section\;\ref{subsec:tradeoff}, $\mu$ was set to $0.01$.

We first visually compare the performance of GGLR with SDGLR. 
As shown in Fig.\;\ref{fig:visualResult1}, interpolation of SDGLR appeared over-smoothed and inconsistent with surrounding textures. 
In contrast, GGLR better preserved image gradients, resulting in more natural reconstruction that was consistent with surrounding textures.  

\begin{table}
\caption{\footnotesize Interpolation results by different methods on 50 images with $\geq 90\%$ randomly missing pixels. GGLR$^4$ (GGLR$^2$) uses 4-connected (2-connected) horizontal and vertical gradient graphs.}
\label{tab:numResult1}
\vspace{-0.1in}
\tiny
\begin{center}
\begin{tabular}{|l|c|c|c|c|c|c|c|c|}
\hline
 \multicolumn{9}{|c|}{ 90\% missing pixels } \\ 
\hline
Methods  &  SDGLR&CSC&IRCNN&IDBP &TGV& GSC&GGLR$^4$&GGLR$^2$\\
\hline
PSNR(dB)   &  21.55 & 20.92 & 22.03& \textbf{22.60} & 22.16 &22.57  &  22.23&21.90\\
\hline
SSIM   & 0.628  & 0.606 &0.651 & 0.680& 0.667&\textbf{ 0.685} & 0.683 &0.670 \\
\hline
 \multicolumn{9}{|c|}{ 95\% missing pixels  } \\ 
\hline
Methods  &  SDGLR&CSC&IRCNN&IDBP &TGV& GSC&GGLR$^4$&GGLR$^2$ \\
\hline
PSNR(dB)   & 19.87   &   18.41 &   16.75  & \textbf{ 20.61}& 20.42 & 20.38 &   20.46&20.24\\
\hline
SSIM   & 0.529  &  0.486 &   0.477  & 0.584  & 0.571& 0.570& \textbf{0.587}&0.575\\
\hline
 \multicolumn{9}{|c|}{  98\% missing pixels } \\ 
\hline
Methods  & SDGLR&CSC&IRCNN&IDBP &TGV& GSC&GGLR$^4$&GGLR$^2$ \\
\hline
PSNR(dB)   & 18.02    &   12.21 & 9.95   & 18.11 & 18.32 &18.09  &\textbf{18.37} &18.17\\
\hline
SSIM  &  0.435  &  0.263  &0.229   & 0.479& 0.477 & 0.456 &\textbf{0.485}&0.471 \\
\hline
 \multicolumn{9}{|c|}{ 99\% missing pixels  } \\ 
\hline
Methods &  SDGLR&CSC&IRCNN&IDBP &TGV & GSC&GGLR$^4$&GGLR$^2$ \\
\hline
PSNR(dB)   &  16.89   &  9.01  &  7.60&16.54  & 17.07&16.85  & \textbf{17.11} &16.90\\
\hline
SSIM  &   0.386 &  0.137   &  0.118   &0.416  & 0.420&0.393 &\textbf{0.423}&0.411 \\
\hline
\end{tabular}
\end{center}
\end{table}

\begin{table}
\caption{\footnotesize Average runtime (in sec) on $128\times 128$ images.}
\label{tab:numResult2}
\scriptsize
\begin{center}
\begin{tabular}{|c|c|c|c|c|c|c|}
\hline
Methods&CSC&GSC&IDBP&TGV &GGLR$^4$& GGLR$^2$\\
\hline
Time (sec.) & $>60$  & $>60$ & 9.58& 9.27  & 2.03 & \textbf{1.04} \\
\hline
\end{tabular}
\end{center}
\end{table}

We compare against state-of-the-art image interpolation methods: SDGLR, TGV \cite{bredies2015tgv}, CSC \cite{zisselman19}, IRCNN \cite{Zhang2017}, IDBP \cite{tirer19} and GSC \cite{Zha2020}, where IRCNN is a DL based method. 
The first 50 images from Weizmann Horse Database \cite{Borenstein02} were used for our experiments.  
The average PSNR and SSIM of the reconstructed images are listed in Table\;\ref{tab:numResult1}. 
We observe that GGLR outperformed competing methods in general when the fraction of missing pixels was large. 
A typical visual comparison is shown in Fig.\;\ref{fig:visualResult2}. 
GGLR preserved image contours and mitigated blocking effects, observed in interpolation by sparse representation (e) and (g).
GGLR achieved the highest PSNR and SSIM when the fraction of missing pixels is above 98\%.

Table\;\ref{tab:numResult2} reports runtime of different methods for a $128\times 128$ image. 
All experiments were run under the Matlab2015b environment on a laptop with Intel Core i5-8365U CPU of 1.60GHz.
TGV employed a primal-dual splitting method \cite{Condat2013} for $\ell_2$-$\ell_1$ norm minimization, which required a large number of iterations until convergence, especially when the fraction of missing pixels was large. 
In contrast, GGLR employing line graphs required roughly $1$sec. 

\vspace{-0.1in}
\section{Conclusion}
\label{sec:conclude}
\vspace{-0.1in}
We propose a high-order signal-dependent gradient graph Laplacian regularizer (GGLR) that promotes piecewise planar (PWP) signal reconstruction in 2D images. 
Unlike total generalized variation (TGV), GGLR is in quadratic form, so that in each iteration, when edge weights are fixed, an unconstrained quadratic programming (QP) problem can be solved efficiently using conjugate gradient (CG). 
Image interpolation experiments show that GGLR outperformed competitors when fraction of missing pixels was large.


\bibliographystyle{IEEEbib}
\bibliography{ref2}

\end{document}